\documentclass{aa}
\usepackage{graphicx}
\usepackage{txfonts}

\usepackage{natbib}
\bibpunct{(}{)}{;}{a}{}{,}

\newcommand{\msun}{\ensuremath{M_\odot}}                          
\newcommand{\rsun}{\ensuremath{R_\odot}}                          
\newcommand{\Msun}{\ensuremath{\,{\rm M}_\odot}}                  
\newcommand{\Rsun}{\ensuremath{\,{\rm R}_\odot}}                  
\newcommand{\lsun}{\ensuremath{L_\odot}}                          
\newcommand{\Teff}{\ensuremath{T_{\rm eff}}}                      
\newcommand{\Vsync}{\ensuremath{V_{\rm synch}}}                   
\newcommand{\EBV}{\ensuremath{E_{B-V}}}                           
\newcommand{\kms}{\,km\,s$^{-1}$}                                 
\newcommand{\mc}[1]{\multicolumn{2}{c}{#1}}                       
\newcommand{\Mbol}{\ensuremath{M_{\rm bol}}}                      
\newcommand{\sci}[2]{{:1}\!\times10^{:2}}                         
\newcommand{\baur}{$\beta$\,Aur}                                  
\newcommand{\wire}{{\sc wire}}                                    
\newcommand{\hms}[3]{\ensuremath{#1^{\rm h}\,#2^{\rm m}\,#3^{\rm s}}}
\newcommand{\dms}[3]{\ensuremath{#1^\circ\,#2'\,#3''}}
\newcommand{\reff}[1]{\textrm{#1}}                                

\begin{document}

\title{Eclipsing binaries observed with the WIRE satellite.}

\subtitle{II. $\beta$\,Aurigae and non-linear limb darkening in light curves
\thanks{Table 2 is only available in electronic form at the CDS via anonymous ftp to {\tt cdsarc.u-strasbg.fr} (130.79.128.5) or via {\tt http://cdsweb.u-strasbg.fr/cgi-bin/qcat?J/A+A/}}}

\titlerunning{WIRE satellite photometry of $\beta$\,Aurigae}

\author{John Southworth\inst{1} \and H.\ Bruntt\inst{2} \and D.\ L.\ Buzasi\inst{3}}

\authorrunning{Southworth, Bruntt \& Buzasi}

\offprints{John Southworth}

\institute{{Department of Physics, University of Warwick, Coventry, CV4 7AL} \email{jkt@astro.keele.ac.uk} \\
           {School of Physics A28, University of Sydney, 2006 NSW, Australia} \email{bruntt@physics.usyd.edu.au} \\
           {US Air Force Academy, Department of Physics, CO, USA} \email{Derek.Buzasi@usafa.af.mil}}
\date{Received ???? / Accepted ????}

\abstract%
{}
{We present the most precise light curve ever obtained of a detached eclipsing binary star and use it investigate the inclusion of non-linear limb darkening laws in light curve models of eclipsing binaries. This light curve, of the bright eclipsing system $\beta$\,Aurigae, was obtained using the star tracker aboard the {{\sc wire}} satellite and contains 30\,000 datapoints with a point-to-point scatter of 0.3\,mmag.}
{We analyse the {{\sc wire}} light curve using a version of the {{\sc ebop}} code modified to include non-linear limb darkening laws and to directly incorporate observed times of minimum light and spectroscopic light ratios into the photometric solution as individual observations. We also analyse the dataset with the Wilson-Devinney code to ensure that the two models give consistent results.}
{{{\sc ebop}} is able to provide an excellent fit to the high-precision {{\sc wire}} data. Whilst the fractional radii of the stars are only defined to a precision of 5\% by this light curve, including an accurate published spectroscopic light ratio improves this dramatically to 0.5\%. Using non-linear limb darkening improves the quality of the fit significantly compared to the linear law and causes the measured radii to increase by 0.4\%. It is possible to derive all of the limb darkening coefficients from the light curve, although they are strongly correlated with each other. The fitted coefficients agree with theoretical predictions to within their fairly large error estimates.
We were able to obtain a reasonably good fit to the data using the Wilson-Devinney code, but only using the highest available integration accuracy and by iterating for a long time. Bolometric albedos of 0.6 were found, which are appropriate to convective rather than radiative envelopes.}
{The radii and masses of the components of \baur\ are $R_{\rm A} = 2.762 \pm 0.017$\Rsun, $R_{\rm B} = 2.568 \pm 0.017$\Rsun, $M_{\rm A} = 2.376 \pm 0.027$\Msun\ and $M_{\rm B} = 2.291 \pm 0.027$\Msun, where A and B denote the primary and secondary star, respectively. Theoretical stellar evolutionary models can match these parameters for a solar metal abundance and an age of 450--500\,Myr. The Hipparcos trigonometric parallax and an interferometrically-derived orbital parallax give distances to \baur\ which are in excellent agreement with each other and with distances derived using surface brightness relations and several sets of empirical and theoretical bolometric corrections.}


\maketitle

\section{Introduction}

The study of detached eclipsing binaries (dEBs) is our primary source of accurate physical properties of normal stars \citep{Andersen91aar}. By modelling their light curves and deriving radial velocities of the two components from their spectra, it is possible to measure the masses and radii of the two components to accuracies of better than 1\% \citep[e.g.,][]{Me+05mn,Lacy+05aj}. The effective temperatures and luminosities of the stars can be found by a variety of different methods, including photometric calibration, spectral analysis and measuring trigonometric parallaxes. These data can be used to measure accurate distances to stellar systems \citep{Ribas+05apj,Me++05aa} and to critically investigate the accuracy of the predictions resulting from theoretical stellar models \citep[e.g., ][]{Ribas06apss}.

We are undertaking a project to obtain accurate and extensive light curves of selected bright dEBs using the Wide Field Infra-red Explorer (\wire) satellite. \wire\ was launched in March 1999 but the coolant for the main camera was lost only a few days after launch. Since then the star tracker on \wire\ has been used to monitor
the variability of bright stars of all spectral types, as described in the review by \citet{Bruntt07}.
The main aims of the current work are to accurately measure the physical properties of the systems and to investigate whether the currently available eclipsing binary light curve models can provide a good fit to observational data with measurement uncertainties substantially below 1\,mmag. We expect the answer to the second question to be `yes', but this matter is becoming increasingly important in light of current and forthcoming space missions, such as Kepler \citep{Basri++05newar} and CoRoT \citep{Michel+06mm}, which aim to obtain extremely high-quality light curves of a large number of stars.

Our first paper in this series presented the serendipitous discovery that the bright star $\psi$\,Centauri is a long-period dEB \citep[][hereafter Paper\,I]{Bruntt+06aa}. The \wire\ light curve of this object shows deep total eclipses arising from a system containing two stars of quite different sizes. dEBs containing two dissimilar stars are particularly useful because the co-evolutionary nature of the two components can put strong constraints on the predictions of theoretical models \citep{Me++04mn2}. However, one of the most interesting aspects of this work is that we were able to measure the relative radii of the two stars (radii expressed as a fraction of the orbital size) to a precision of better than 0.2\%, where the uncertainties were estimated from extensive Monte Carlo simulations but did not include error contributions from some sources such as the use of different algebraic laws to describe the limb darkening of the stars.

\subsection*{\baur igae}

In this work we present a \wire\ light curve for the dEB \baur igae (Menkalinan), which is composed of two A1\,IV stars orbiting each other with a period of 3.96 days (Table\,\ref{table:betaur}). This light curve contains 30\,015 datapoints with a point to point scatter of less than 0.3\,mmag. We chose \baur\ as a target for the {\sc wire} satellite due to its brightness, which not only made our observations more precise but also makes it a very difficult object to study with modern telescopes. In addition, it has an accurate trigonometric parallax from the {\it Hipparcos} satellite which means the effective temperatures and surface brightnesses of the stars can be measured directly and used to calibrate these quantities for other objects.

\baur\ was one of the first stars discovered to be a spectroscopic binary system, by \citet{Maury90}, and an orbit was calculated by \citet{Rambaut91mn} from the published observations. It was subsequently discovered by \citet{Stebbins11apj} to undergo shallow eclipses, and so was also one of the first known eclipsing binaries. A detailed discussion of early work, and an investigation into whether the speed of light is dependent on wavelength, was given by \citet{Baker10}.


The brightness of \baur\ and the shallowness of its eclipses has made it difficult to observe photometrically, and only the study by \citet{Johansen71aa} has provided accurate results. Johansen obtained light curves of \baur\ in seven passbands, four of which closely resemble the Str\"omgren $uvby$ system, and solved them using the method of \citet{RussellMerrill52book}.

However, the brightness of \baur\ and the relatively low rotational velocities of the two components means that it is observationally one of the most straightforward spectroscopic binaries to study. Spectroscopic radial velocities have been presented by \citet{Rambaut91mn}, \citet{Vogel04apj}, \citet{Smith48apj}, \citet{GaleottiGuerrero68ibvs} and \citet{PopperCarlos70pasp}, the most accurate and detailed being that by Smith. In addition to these works, the spectrum of \baur\ has been considered by \citet{Maury98apj}, \citet{StruveDriscoll53pasp} and \citet{Toy69apj}, the most recent study \citep{Lyubimkov++96arep} demonstrating that the two components have almost identical spectra and exhibit enhanced metallic lines typical of the Am class of stars.

\citet{Hummel+95aj} have interferometrically resolved the orbit of \baur\ using the Mark\,III Optical Interferometer. This means that the distance to the system can be obtained straightforwardly by comparing their angular orbital semimajor axis to the linear value measured using spectroscopy. In addition, Hummel et al.\ have measured the orbital inclination, which can be compared with the value derived from the light curve to ensure consistency.

\citet[][hereafter NJ94]{NordstromJohansen94aa} derived accurate absolute dimensions of \baur\ by analysing the light curves of \citet{Johansen71aa} with the {\sc ebop} code and adopting the spectroscopic orbits from \citet{Smith48apj}. It is well known that for dEBs containing two stars of almost the same size, particularly if they exhibit shallow eclipses, the ratio of the stellar radii is poorly determined by the light curves \citep{Popper84aj}, so NJ94 constrained their light curve solution with an accurate spectroscopic light ratio. The resulting properties of the stars have since been used for studies of the helium--metal enrichment law \citep{Ribas+00mn} and the fundamental effective temperature scale \citep{Smalley+02aa}.

\begin{table} \caption{Identifications and astrophysical data for \baur.
\newline {\bf References:} (1) \citet{HoffleitJaschek91}; (2) \citet{CannonPickering18};
(3) \citet{Perryman+97aa}; (4) NJ94; (5) \citet{Lyubimkov++96arep}; (6) \citet{Hog+97aa}.}
\label{table:betaur} \centering
\begin{tabular}{lr@{}lc} \hline \hline
                              &     & \baur                 & References\\ \hline
Bright Star Catalogue         &     & HR 2088               & 1       \\
Henry Draper Catalogue        &     & HD 40183              & 2       \\
Hipparcos Catalogue           &     & HIP 28360             & 3       \\
$\alpha_{2000}$               &     & \hms{05}{59}{31.7229} & 3       \\
$\delta_{2000}$               & $+$ & \dms{44}{56}{50.758}  & 3       \\[2pt]
Spectral type                 &     & A1m\,IV + A1m\,IV     & 4,5     \\
Trigonometric parallax        &     & 39.72 $\pm$ 0.78 $m$as& 3       \\
Distance                      &     & 25.18 $\pm$ 0.49 pc   & 3       \\
Tycho $B$ magnitude ($B_T$)   &     & 1.989 $\pm$ 0.005     & 6       \\
Tycho $V$ magnitude ($V_T$)   &     & 1.904 $\pm$ 0.005     & 6       \\
\hline \end{tabular} \end{table}


\section{Observations and data reduction}

\baur\ was observed with the \wire\ star tracker from 2006 March 30 to April 20. During each 93.2\,min orbit of Earth \wire\ switched between two targets in order to minimise the influence of scattered light from the daytime side of Earth. In the data reduction we discarded the datapoints at the very beginning and the end of each orbit, resulting in a duty cycle of 25\%. Observations were obtained during every orbit for 21 days. The \wire\ passband is approximately $V$+$R$ \citep{Bruntt+07aa}. \reff{It may in future be possible to obtain a more precise description of the response function by considering the the total set of observations obtained using the \wire\ star tracker.}

\begin{figure*} \centering \includegraphics[width=\textwidth]{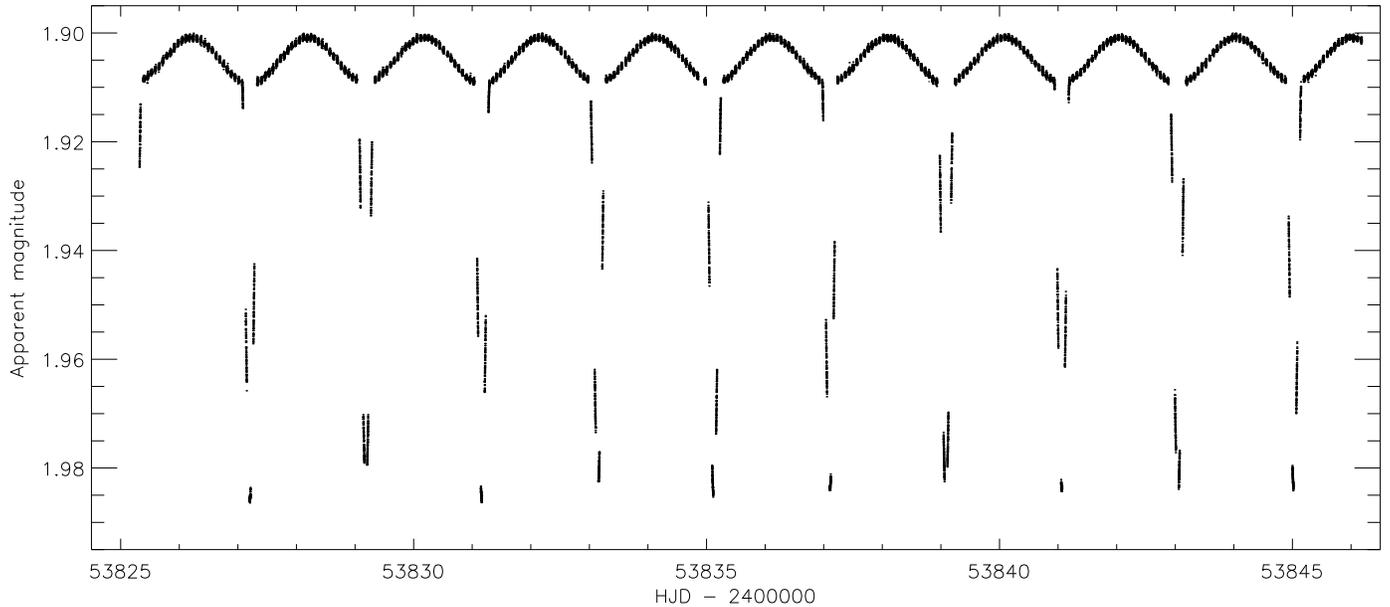}
\caption{The \wire\ light curve of \baur. Each group of points represents
continuous observation for a part of one orbit of the \wire\ satellite
around Earth. The overall magnitude has been adjusted to match the Hipparcos
$H_p$ light curve but is not exact because the passband of \wire\ is
unknown.} \label{fig:lcplot} \end{figure*}

The raw data comprises about 950\,000 CCD windows of $8 \times 8$ pixels collected at a rate of 2\,Hz. To extract the light curve we used the photometric pipeline described by \citet{Bruntt+05apj}. The background level was very low in these observations (20\,ADU or 300 $e^-$), and hence the photometric precision is mainly determined from Poisson statistics. We binned the light curve to four datapoints per minute, giving a total of 30\,015 datapoints after rejecting a few clear outliers.

The point-to-point scatter in this light curve is 0.28\,mmag, of which the Poisson noise accounts for 0.21\,mmag as each original observation contains about 770\,000 $e^-$. We calculated the noise in the Fourier amplitude spectrum in the frequency range 2--3\,mHz. The mean amplitude level is $\langle A \rangle=2.7$\,$\mu$mag, which corresponds to $\sqrt{\langle A \rangle * N / \pi} = 0.27$\,mmag per data point. The observed noise level for \baur\ is higher by 29\% than the Poisson noise. \citet{Bruntt+05apj} analysed \wire\ data of Procyon and their measured noise level was 15--20\% higher than the Poisson noise. These higher noise levels may be due to instrumental effects like satellite jitter combined with variable sub-pixel efficiency in the CCD.

After subtracting a preliminary light curve fit we clearly saw long-period drifts at low amplitude. In particular, for the first day of observations the star appears to slowly decrease in brightness by 2\,mmag and during the last two days of observations the star apparently again became brighter by 2\,mmag. This change is tightly correlated with a slight increase in the background level, and is clearly not intrinsic to \baur\ as it is not affected by the presence of eclipses. In the Fourier spectrum of the residual light curve we detected two peaks at $f_1 = 1.020$ and $f_2 = 5.148$ cycle\,d$^{-1}$, both with amplitudes of just 70\,$\mu$mag. The former is likely related to a slight change of the phase of the illuminated Earth. The second frequency corresponds to exactly one third of the orbital frequency of WIRE ($f_{\rm WIRE} = 15.449$\,cycle\,d$^{-1}$). We cannot be sure why this is visible in the data. To remove the effects mentioned above from the light curve, we used a high-pass filter on the observations. This was done on the {\em residuals} of the preliminary light curve fit, and four iterations were made to ensure the depths or shapes of the eclipses were not affected. The high-pass filter has a cut-off which removes drifts in the data corresponding to periods longer than 24 hours.

We note that in the analysis of WIRE data of the binary $\psi$\,Cen (Paper\,I) we found two significant peaks at $f_1 = 1.996(2)$ and $f_2 = 5.127(3)$ cycle\,d$^{-1}$, both with amplitudes around 0.2\,mmag. In Paper\,I we proposed that these frequencies were due to pulsations intrinsic to the star. However, at the time of those observations the orbital frequency of WIRE was slightly lower at $f_{\rm WIRE} = 15.348 \pm 0.001$\,cycle\,d$^{-1}$, which is exactly one third of $f_2$ as we also find here for \baur. It is therefore very likely that the variation found at the frequency $\frac{1}{3}f_{\rm WIRE}$ is {\em not} due to variations intrinsic to either $\psi$\,Cen or \baur.


\section{Light curve analysis}

\begin{table} \caption{\wire\ differential-magnitude light curve for \baur.
The complete table is available in the electronic version of this work.}
\label{tab:wirelc} \centering
\begin{tabular}{lc} \hline \hline
HJD & Differential magnitude \\ \hline
2453825.314879 & 0.0195836 \\
2453825.315059 & 0.0197159 \\
2453825.315238 & 0.0193520 \\
2453825.315417 & 0.0190368 \\
2453825.315597 & 0.0192776 \\
2453825.315776 & 0.0186899 \\
2453825.315956 & 0.0187692 \\
2453825.316135 & 0.0190573 \\
2453825.316314 & 0.0187044 \\
2453825.316494 & 0.0184744 \\
\hline \end{tabular} \end{table}

\begin{figure*} \centering \includegraphics[width=\textwidth]{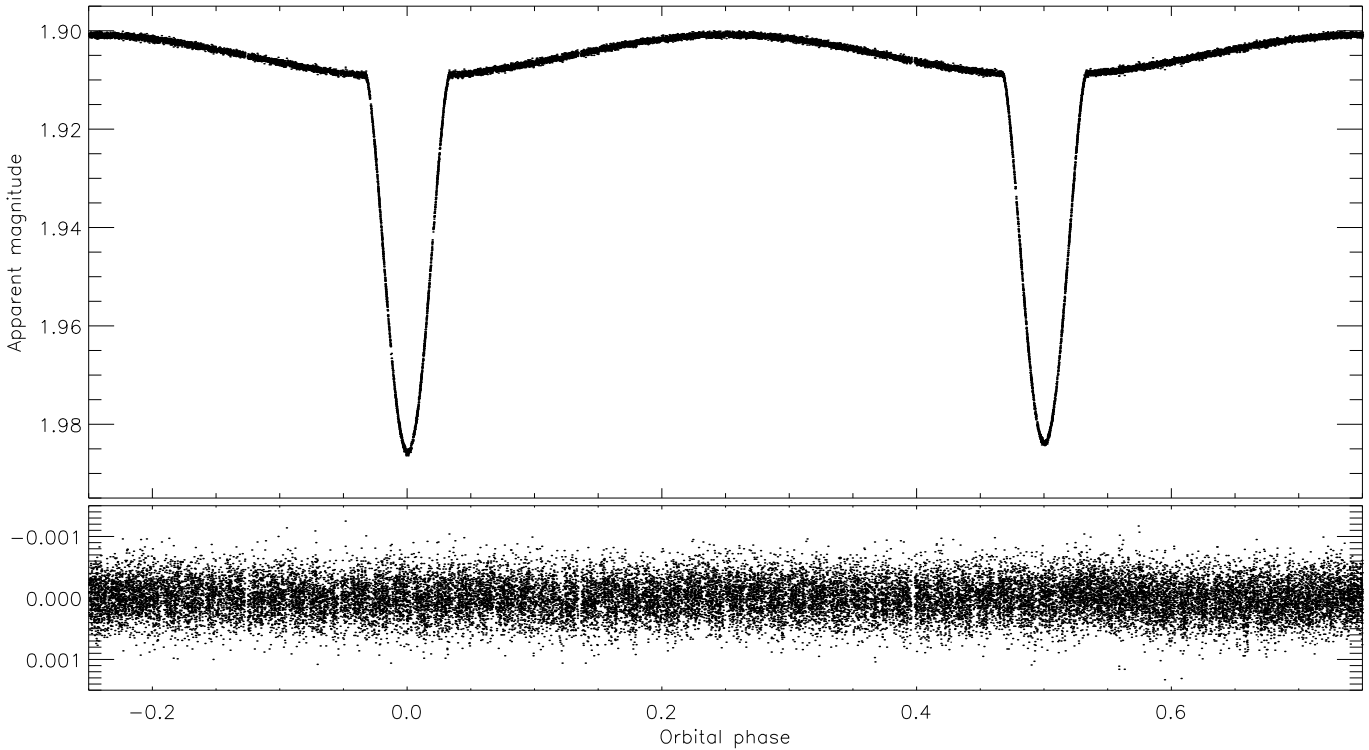}
\caption{Phased \wire\ light curve of \baur. The residuals of the best
fit are shown in the lower panel. The best fit has not been plotted in the
upper panel as it completely overlaps the data (as can be seen from the
residuals).} \label{fig:lcfit} \end{figure*}

The \wire\ light curve of \baur\ is shown in Fig.\,\ref{fig:lcplot}. It lasts for just under 21 days and contains 11 eclipses. The observations are presented in Table\,\ref{tab:wirelc} (available in its entirety in the electronic version of this work). We define phase zero to occur at the midpoint of the primary eclipse, which is deeper than the secondary eclipse. We will refer to the eclipsed component at this point as star A and the (cooler) eclipsing body as star B. For \baur\ star A is slightly larger and more massive than star B, as well as being a little hotter. The phased \wire\ light curve is shown in Fig.\,\ref{fig:lcfit} with the residuals of the best fitting model (see below).

For the main analysis of the \wire\ light curve we have used {\sc jktebop}\footnote{{\sc jktebop} is written in {\sc fortran77} and the source code is available at {\tt http://www.astro.keele.ac.uk/$\sim$jkt/}} \citep{Me++04mn2}, which is based on the {\sc ebop} code \citep{PopperEtzel81aj,Etzel81conf}, which in turn uses the NDE biaxial ellipsoidal model for well-detached eclipsing binaries \citep{NelsonDavis72apj,Etzel75}. {\sc jktebop} incorporates several modifications to the standard code, particularly in input/output and the use of the Levenberg-Marquardt \reff{minimisation} algorithm ({\sc mrqmin}; see \citealt{Press+92book}, p.\,678) for finding the optimal model parameters for describing a dataset. The relative radii of the stars are defined to be $r_{\rm A} = \frac{R_{\rm A}}{a}$ and $r_{\rm B} = \frac{R_{\rm B}}{a}$, where $R_{\rm A}$ and $R_{\rm B}$ are the absolute radii of the stars and $a$ is the orbital semimajor axis. {\sc jktebop} solves for the sum and ratio of the relative radii \citep{Me+05mn} as these are more directly related to the morphology of dEB light curves than the individual radii.

For this study we have implemented three additional modifications: the incorporation of non-linear limb darkening and the direct inclusion of spectroscopic light ratios and times of minimum light as observational data. These will be discussed below.

\subsection{Period analysis}                                                                   \label{sec:lc:tmin}

A common method of determining the linear orbital ephemeris of an eclipsing binary is to collect times of minimum light, spanning as large a time interval as possible, and fit them with a straight line against cycle number \citep[e.g.,][]{Me+05mn}. This procedure is not a good approach here due to the properties of the \wire\ light curve: it is possible to obtain individual times of minimum by fitting Gaussian functions separately to each eclipse, but these contain systematic errors in the measured times as the eclipse shape is slightly non-Gaussian and individual eclipses are poorly sampled. An improvement over this would be to fit each eclipse with a light curve model, but the best way to measure an orbital period for an eclipsing binary is to fit for it directly using the whole light curve. This is often not possible because the available times of minimum light are not accompanied by the data from which they were measured.

We have therefore modified {\sc jktebop} to treat times of minimum light themselves as observational data and include these alongside the light curve when finding the best fit. The inclusion of disparate types of data obviously requires the measurement errors of each type to be accurately determined so their relative weightings are correct \citep[see][]{Wilson93aspc}. We have measured times of minimum light from the \wire\ light curve using Gaussian functions. We report these for the interested reader (Table\,\ref{tab:tmin}) but have not used them in our analysis.

A detailed period analysis for \baur\ was given by \citet{JohansenSorensen97ibvs} on the basis of several observed times of minimum light and spectroscopically or interferometrically determined times of conjunction. Whilst there have been occasional suggestions that the period of \baur\ is not constant \citep[see][]{Vogel04apj}, Johansen \& S{\o}rensen found no evidence for a variable orbital period. We have chosen to include only those times of minimum light which were obtained via photoelectric or CCD photometry as these measurements are more accurate than the times of conjunction obtained in other ways.

In order to derive the orbital period of \baur\ without accidentally introducing cycle count errors we began with a preliminary light curve fit to the the \wire\ observations. Once we had determined the best period for this dataset we added in the Hipparcos light curve \citep{Perryman+97aa}, with appropriate observational errors, to give a more accurate period (see also Paper\,I). The slightly different bandpasses for these two datasets is not important because the quality and quantity of the \wire\ observations are both orders of magnitude better than the Hipparcos data. Inclusion of the latter makes a negligible difference to the parameters of the fit, except for increasing the accuracy of the orbital period determination.

We then introduced the times of minimum light from \citet{JohansenSorensen97ibvs} and \citet{Johansen71aa} before finally extending the dataset to cover almost a century with the time of minimum from \citet{Stebbins11apj} quoted by Johansen \& S{\o}rensen. The quality of the fit to the times of minimum is excellent (Table\,\ref{tab:tmin}), and from consideration of these data and the other times of minimum used by Johansen \& S{\o}rensen we can find no evidence for a change in period.

\begin{table*} \caption{Times of minimum light for \baur\ compared to the calculated
values for our chosen model fit. Times are given in HJD $-$ 2\,400\,000.
\newline $^*$ Note that the times measured from the \wire\ data are expected to be
affected by systematic errors (see Section\,\ref{sec:lc:tmin}) and were not included
in the fit. They are repeated here for completeness and for the interested reader.}
\label{tab:tmin} \centering
\begin{tabular}{r r@{.}l@{\,$\pm$\,}l c c l} \hline \hline
Orbital cycle   &  \multicolumn{3}{c}{Observed time} & Calculated time & $O-C$ & Reference \\
\hline
$-$8790.0       & 19018 & 3848  & 0.0046  & 19018.38497       & $-$0.00017 & \citet{Stebbins11apj} \\
$-$4289.0       & 36842 & 5537  & 0.0013  & 36842.55528       & $-$0.00158 & \citet{Johansen71aa} \\
$-$4085.0       & 37650 & 4025  & 0.0011  & 37650.40481       & $-$0.00231 & \citet{Johansen71aa} \\
$-$3719.0       & 39099 & 785   & 0.0011  & 39099.78191       & $+$0.00309 & \citet{Johansen71aa} \\
$-$1030.0       & 49748 & 353   & 0.0023  & 49748.34755       & $+$0.00545 & \citet{JohansenSorensen97ibvs} \\
 $-$855.0       & 50441 & 3552  & 0.0035  & 50441.35573       & $-$0.00053 & \citet{JohansenSorensen97ibvs} \\[2pt]
      0.0\,$^*$ & 53827 & 19586 & 0.00045 & 53827.19567\,$^*$ & $+$0.00019 & This work \\
      0.5\,$^*$ & 53829 & 17566 & 0.00047 & 53829.17570\,$^*$ & $-$0.00004 & This work \\
      1.0\,$^*$ & 53831 & 15541 & 0.00054 & 53831.15572\,$^*$ & $-$0.00031 & This work \\
      1.5\,$^*$ & 53833 & 13593 & 0.00029 & 53833.13574\,$^*$ & $+$0.00019 & This work \\
      2.0\,$^*$ & 53835 & 11562 & 0.00032 & 53835.11577\,$^*$ & $-$0.00015 & This work \\
      2.5\,$^*$ & 53837 & 09603 & 0.00042 & 53837.09579\,$^*$ & $+$0.00024 & This work \\
      3.0\,$^*$ & 53839 & 07576 & 0.00019 & 53839.07581\,$^*$ & $-$0.00005 & This work \\
      3.5\,$^*$ & 53841 & 05571 & 0.00047 & 53841.05584\,$^*$ & $-$0.00013 & This work \\
      4.0\,$^*$ & 53843 & 03598 & 0.00024 & 53843.03586\,$^*$ & $+$0.00012 & This work \\
      4.5\,$^*$ & 53845 & 01563 & 0.00037 & 53845.01588\,$^*$ & $-$0.00025 & This work \\
\hline \end{tabular} \end{table*}

\subsection{Direct inclusion of a spectroscopic light ratio in the light curve fitting process}

For dEBs which contain two similar stars undergoing partial eclipses, the sum of the radii ($r_{\rm A}+r_{\rm B}$) is well defined as it depends mainly on the orbital inclination and duration of the eclipses. The ratio of the radii ($k = \frac{r_{\rm B}}{r_{\rm A}}$) can be very poorly defined by the observed light curves of such systems because quite different values can give very similar eclipse shapes \citep[e.g.][]{Andersen++80iaus}. \baur\ is unfortunately an excellent example of this situation, despite the quantity and quality of the \wire\ light curve. However, the light ratio of dEBs is normally very strongly dependent on $k$ so an observed spectroscopic light ratio can be used to constrain $k$ and therefore the individual stellar radii \citep[e.g.][]{Me++04mn,MeClausen07aa}.

The usual way this constraint is applied is to map out how the light ratio depends on $k$ and use the independent spectroscopic light ratio to specify the allowable values of $k$ \citep[e.g.][]{Torres+00aj}. However, a more rigorous procedure is to include the spectroscopic light ratio directly as an observation to be used during the minimisation process. This also greatly simplifies the error analysis. {\sc jktebop} has been modified to explicitly include this type of observable quantity in the fitting process. As with inclusion of the times of minimum light, it is important to ensure that accurate measurement errors are adopted when mixing different types of data in one fitting process.

In preliminary fits to the \wire\ light curve we found that $r_{\rm A}$ and $r_{\rm B}$ were determined to a precision of only 5\% (uncertainty from Monte Carlo simulations; see below). This was worse than expected, and quite distant from the target of 2\% which is the minimum for obtaining useful results for most dEBs \citep{Andersen91aar,Andersen98iaus,Gimenez92iaus}. To solve this problem in their analysis of \baur, NJ94 obtained a high-quality spectroscopic light ratio of $0.855 \pm 0.016$, which they used to constrain $k$ and so obtain accurate values for $r_{\rm A}$ and $r_{\rm B}$. This light ratio is consistent with and much more accurate than the magnitude differences of $\Delta m = 0.2 \pm 0.2$\,mag given by \citet{Hummel+95aj} and of $\Delta m = 0.13$\,mag given by \citet{Petrie39pdao}. When we included the NJ94 light ratio in our fit we found that the precision of the radii improved dramatically from 5\% to 0.5\%. $k$ is determined almost entirely by the light ratio in this solution.

A possible problem in this approach is that NJ94's light ratio was obtained using the Mg\,II spectral line at 4481\,\AA\ but the {\sc wire} passband is redder than this so the light ratio may be different. However, the two stars have very similar effective temperatures so the systematic error caused by this effect should be negligible. We note that it is not possible to prove this because the transmission curve of the {\sc wire} star tracker is not precisely known. In the future we will obtain a spectroscopic light ratio at a wavelength closer to the effective wavelength of the {\sc wire} passband.

\begin{table*} \caption{Best-fitting photometric parameters. These solution have been obtained
by fitting the \wire\ and Hipparcos light curve as well as the spectroscopic light ratio
of NJ94 and several times of minimum light. Times are expressed in HJD $-$ 2\,400\,000.}
\label{tab:lcfit} \centering
\begin{tabular}{l l r@{\,$\pm$\,}l r@{\,$\pm$\,}l r@{\,$\pm$\,}l} \hline \hline
\multicolumn{8}{l}{\it Results using individual LD coefficients for each star} \\
                           &                   &   \mc{Linear LD law}    & \mc{Square-root LD law} &  \mc{Quadratic LD law}  \\
\hline
Sum of the relative radii  & $r_{\rm A}+r_{\rm B}$ &   0.30174   & 0.00007   &   0.30297   & 0.00036   &   0.30290   & 0.00047   \\
Ratio of the radii         & $k$               &   0.9289    & 0.0083    &   0.9311    & 0.0083    &   0.9300    & 0.0084    \\
Surface brightness ratio   & $J$               &   0.9883    & 0.0028    &   0.9812    & 0.0091    &   0.9856    & 0.0126    \\
Orbital inclination (deg.) & $i$               &  76.862     & 0.004     &  76.809     & 0.015     &  76.803     & 0.020     \\
Eccentricity term          & $e\cos\omega$     &$-$0.000017  & 0.000006  &$-$0.000016  & 0.000006  &$-$0.000015  & 0.000006  \\
Eccentricity term          & $e\sin\omega$     &   0.00191   & 0.00032   &   0.00173   & 0.00036   &   0.00183   & 0.00035   \\
Orbital period (days)      & $P_{\rm orb}$     &   3.96004673& 0.00000015&   3.96004673& 0.00000015&   3.96004673& 0.00000015\\
Time of primary minimum    & $T_{\rm Min\,I}$  &52827.195695 & 0.000011  &52827.195693 & 0.000011  &52827.195693 & 0.000011  \\
Linear LD coefficient      & $u_{\rm A}$       &   0.382     & 0.006     &$-$0.056     & 0.147     &   0.210     & 0.064     \\
Linear LD coefficient      & $u_{\rm B}$       &   0.379     & 0.006     &$-$0.177     & 0.148     &   0.187     & 0.060     \\
Non-linear LD coefficient  & $v_{\rm A}$       &         \mc{\ }         &   0.703     & 0.232     &   0.248     & 0.096     \\
Non-linear LD coefficient  & $v_{\rm B}$       &         \mc{\ }         &   0.881     & 0.236     &   0.277     & 0.094     \\
\hline
Relative radius of star A  & $r_{\rm A}$       &   0.15643   & 0.00065   &   0.15689   & 0.00068   &   0.15694   & 0.00070   \\
Relative radius of star B  & $r_{\rm B}$       &   0.14531   & 0.00069   &   0.14608   & 0.00072   &   0.14595   & 0.00074   \\
Light ratio      & $\ell_{\rm B}/\ell_{\rm A}$ &   0.853     & 0.016     &   0.855     & 0.016     &   0.854     & 0.016     \\
Orbital eccentricity       & $e$               &   0.00191   & 0.00032   &   0.00173   & 0.00036   &   0.00183   & 0.00035   \\
Periastron longitude (deg.)& $\omega$          &  90.51      & 0.21      &  90.51      & 0.25      &  90.47      & 0.23      \\
Reduced $\chi^2$        & $\chi^2_{\rm \ red}$ &        \mc{1.332}       &         \mc{1.328}      &        \mc{1.327}        \\
\hline \\ \hline \hline
\multicolumn{8}{l}{\it Results using the same LD coefficients for the two stars} \\
                           &                   &    \mc{Linear LD law}    &  \mc{Square-root LD law} &   \mc{Quadratic LD law}  \\
\hline
Sum of the relative radii  & $r_{\rm A}+r_{\rm B}$ &   0.30174   & 0.00005   &   0.30267   & 0.00039   &   0.30295   & 0.00051   \\
Ratio of the radii         & $k$               &   0.9291    & 0.0081    &   0.9292    & 0.0083    &   0.9295    & 0.0083    \\
Surface brightness ratio   & $J$               &   0.9889    & 0.0027    &   0.9904    & 0.0027    &   0.9908    & 0.0026    \\
Orbital inclination (deg.) & $i$               &   76.862    & 0.012     &  76.821     & 0.016     &  76.800     & 0.022     \\
Eccentricity term          & $e\cos\omega$     &$-$0.000017  & 0.000006  &$-$0.000016  & 0.000006  &$-$0.000015  & 0.000006  \\
Eccentricity term          & $e\sin\omega$     &   0.00178   & 0.00026   &   0.00198   & 0.00027   &   0.00201   & 0.00026   \\
Orbital period (days)      & $P_{\rm orb}$     &   3.96004673& 0.00000015&   3.96004673& 0.00000015&   3.96004673& 0.00000015\\
Time of primary minimum    & $T_{\rm Min\,I}$  &52827.195695 & 0.000011  &52827.195694 & 0.000011  &52827.195693 & 0.000011  \\
Linear LD coefficient      & $u$               &   0.380     & 0.017     &   0.049     & 0.138     &   0.191     & 0.062     \\
Non-linear LD coefficient  & $v$               &         \mc{\ }         &   0.597     & 0.223     &   0.273     & 0.098     \\
\hline
Relative radius of star A  & $r_{\rm A}$       &   0.15642   & 0.00065   &   0.15689   & 0.00066   &   0.15701   & 0.00068   \\
Relative radius of star B  & $r_{\rm B}$       &   0.14533   & 0.00075   &   0.14578   & 0.00073   &   0.14594   & 0.00076   \\
Light ratio       &$\ell_{\rm B}/\ell_{\rm A}$ &   0.853     & 0.016     &   0.854     & 0.016     &   0.855     & 0.016     \\
Orbital eccentricity       & $e$               &   0.00178   & 0.00026   &   0.00198   & 0.00027   &   0.00201   & 0.00026   \\
Periastron longitude (deg.)& $\omega$          &  90.55      & 0.22      &  90.46      & 0.20      &  90.43      & 0.19      \\
Reduced $\chi^2$        & $\chi^2_{\rm \ red}$ &        \mc{1.332}       &         \mc{1.328}      &        \mc{1.327}        \\
\hline \end{tabular} \end{table*}

\begin{table*} \caption{Comparison between the measured LD coefficients for \baur\
and predictions calculated using theoretical model atmospheres. Bilinear interpolation
was used to interpolate the tabulated theoretical coefficients to the effective
temperatures and surface gravities for \baur\ given by NJ94. A solar metal abundance
and a microturbulence velocity of 2\kms\ have been adopted and results are quoted for
the Johnson $V$ and Cousins $R$ bandpasses.
\newline {\bf References}: Van Hamme is \citet{Vanhamme93aj}; D\'{\i}az-Cordoves is
\citet{Diaz++95aas} and \citet{Claret++95aas}; Claret is \citet{Claret00aa} and
Claret \& Hauschildt is \citet{ClaretHauschildt03aa}.}
\label{tab:ld} \centering
\begin{tabular}{l c r@{\,$\pm$\,}l rr rr rr rr rr } \hline \hline
LD coefficient &Star&   \mc{\baur}    & \mc{Van Hamme}  & \mc{D\'{\i}az-Cordoves} & \mc{Claret ({\sc atlas9})} & \mc{Claret ({\sc phoenix})} & \mc{Claret \& Hauschildt} \\
               &    &     \mc{ }      & $V_J$ & $R_C$   & $V_J$  & $R_C$  & $V_J$  & $R_C$  & $V_J$  & $R_C$  & $V_J$  & $R_C$  \\
\hline
Linear law     & A  &  0.382 & 0.006 & 0.437 &   0.367 & 0.525 &   0.418 &  0.536 &   0.542 & 0.534 &   0.454 &   0.489 &   0.406 \\
Linear law     & B  &  0.379 & 0.006 & 0.445 &   0.372 & 0.532 &   0.402 &  0.451 &   0.455 & 0.541 &   0.459 &   0.488 &   0.404 \\
Sqrt (linear)  & A  &$-$0.06 & 0.15  & 0.005 &$-$0.081 & 0.016 &$-$0.019 &  0.035 &$-$0.005 & 0.008 &$-$0.044 &$-$0.245 &$-$0.240 \\
Sqrt (linear)  & B  &$-$0.18 & 0.15  & 0.015 &$-$0.003 & 0.028 &$-$0.017 &  0.046 &   0.001 & 0.017 &$-$0.041 &$-$0.186 &$-$0.183 \\
Sqrt (nonlinear)& A &   0.70 & 0.23  & 0.720 &   0.625 & 0.722 &   0.598 &  0.664 &   0.603 & 0.695 &   0.658 &   1.128 &   0.993 \\
Sqrt (nonlinear)& B &   0.88 & 0.24  & 0.717 &   0.625 & 0.714 &   0.599 &  0.657 &   0.602 & 0.692 &   0.661 &   1.036 &   0.902 \\
Quad (linear)  & A  &   0.21 & 0.06  &       &         & 0.273 &   0.204 &  0.208 &   0.159 & 0.203 &   0.146 &   0.242 &   0.192 \\
Quad (linear)  & B  &   0.19 & 0.06  &       &         & 0.282 &   0.196 &  0.217 &   0.164 & 0.211 &   0.150 &   0.258 &   0.206 \\
Quad (nonlin.) & A  &   0.25 & 0.10  &       &         & 0.328 &   0.279 &  0.387 &   0.345 & 0.401 &   0.372 &   0.366 &   0.318 \\
Quad (nonlin.) & B  &   0.28 & 0.09  &       &         & 0.325 &   0.268 &  0.383 &   0.344 & 0.400 &   0.374 &   0.340 &   0.294 \\
\hline \end{tabular} \end{table*}

\subsubsection{Non-linear limb darkening}                   \label{sec:nld}

A complication in fitting the light curve is the inclusion of limb darkening (LD). The {\sc ebop} code uses the linear LD law \citep{Russell12apj2}, which normally provides an entirely acceptable fit to the light curves of dEBs but is known to be a woefully inadequate representation of the predictions of stellar model atmospheres \citep[e.g.][]{Vanhamme93aj}. There is evidence from studies using the Wilson-Devinney code \citep{WilsonDevinney71apj} that including bi-parametric laws can cause a slight improvement in the fit and a systematic increase of the order of 0.2\% in the measured fractional stellar radii (Paper\,I; \citealt{Lacy+05aj}), which is certainly large enough to be important in the present case. This can be understood because non-linear LD laws generally result in a lower flux towards the stellar limb compared to the linear law \citep[see e.g.][]{Claret00aa,ClaretHauschildt03aa}, so a larger area is required to produce the same amount of light at the limb.

{\sc jktebop} has therefore been modified to allow both the use of several non-linear LD laws\footnote{\reff{Non-linear LD laws have previously been incorporated into {\sc ebop} by \citet{Diaz90phd}}. We are grateful to Dr.\ A.\ Gim\'enez for sending us the relevant lines of code. See \citet{GimenezDiaz93conf} and \citet{GimenezQuintana92aa}.} and the ability to optimise {\em both} coefficients for each star. The implemented laws are the quadratic \citep{Kopal51}, square-root \citep{DiazGimenez92aa} and logarithmic \citep{KlinglesmithSobieski70aj}.

It is common practise when fitting the light curves of eclipsing binaries to use LD coefficients obtained from theoretical model stellar atmospheres, partly because the light curve shape is not strongly dependent on the LD coefficients and a good set of data is needed to be able to fit for them directly \citep{Popper84aj,North++97aa}. However, this means that the final solution depends on theoretical predictions, so it is better to fit for the coefficients when possible.

We have performed light curve solutions for \baur\ using the linear, square-root and quadratic LD laws. In each case we included the LD coefficients for both stars in the set of fitted parameters to avoid using theoretical coefficients. Because the two stars are very similar we also investigated solutions using the same LD coefficients for both stars.

\subsection{Light curve solution}                                                         \label{sec:lc:soln}

When obtaining the best fit to the available data we included the \wire\ light curve (30\,015 datapoints), the Hipparcos $H_p$ light curve (131 datapoints), six photoelectrically-measured times of minimum light and the spectroscopic light ratio obtained by NJ94. The \wire\ observations themselves have a root-mean-squared residual of 0.3\,mmag per point. We found that the orbital eccentricity is small but significant. When third light was included in the photometric solution we found that only a small and ill-defined amount was needed, resulting in very minor changes in other parameters. We have therefore assumed that the light curve contains no contribution in addition to those from the two components of the dEB. A mass ratio of 0.97 was adopted from the spectroscopic study of \citet{Smith48apj}, and gravity brightening exponents were set to 1.0 \citep{Claret98aas}; large changes in these parameters have a negligible effect on the light curve as the stars are well separated in their orbit.

When fitting the data we optimised $r_{\rm A}+r_{\rm B}$, $k$, the LD coefficients, the orbital inclination ($i$), the ratio of the central surface brightnesses of the two stars ($J$), the orbital period ($P$) and the reference time of minimum light ($T_{\rm Min\,I}$). Orbital eccentricity, $e$, and the longitude of periastron, $\omega$, were included in the usual {\sc ebop} way, by fitting for the quantities $e\cos\omega$ and $e\sin\omega$ which have a more direct dependence on the shape of the light curve than $e$ and $\omega$ \citep{Etzel75,PopperEtzel81aj}. The out-of-eclipse magnitude of the system was also included as a fitted parameter. The {\sc wire} data do not have a well-defined apparent magnitude zeropoint so this parameter has only a limited meaning and will not be discussed further.

Our initial attempts to include the spectroscopic light ratio from NJ94 were only partially successful. However, this problem disappeared completely if we included the reflection coefficients as free parameters rather than calculating them analytically from the system geometry. {\sc ebop}'s simple analytical treatment of the reflection effect uses the uniformly-illuminated-hemisphere approach of \citet{Binnendijk74va} and despite providing an apparently good fit to the data it appears to be inadequate for observations of the quality of the \wire\ light curve. This problem can be avoided by including the reflection coefficients as parameters of the fit rather than calculating them from the geometry of the system. Their values are very well determined as they depend on the amplitude of the light variation outside eclipse, and this approach allows us to obtain an excellent  fit to the light curve and spectroscopic light ratio.

Measurement errors for the \wire\ data are not available, so once this solution was obtained we calculated the appropriate measurement errors based on the residuals of the fit. We then binned these data by a factor of six, reducing the number of datapoints to 5027. In Paper\,I we noticed that a similar procedure during the analysis of $\psi$\,Cen resulted in fits which had a reduced chi-squared ($\chi^2_{\rm \ red}$) of 1.15 rather than 1.0, meaning that there was a non-random contribution to the residuals of the fit. We attributed this to a small systematic contribution to the photometric errors in the \wire\ data and multiplied the final uncertainties on each parameter by $\sqrt{\chi^2_{\rm \ red}}$ to take this into account. Here the same effect is present, and the $\chi^2_{\rm \ red}$ values are in the region of 1.33. Minor systematic residuals in the fit might also contribute to $\chi^2_{\rm \ red}$, and in this case are possibly attributable to the approximation used for the reflection effect. It must be remembered that the size of this effect is extremely small and would be essentially undetectable in a dataset with photometric errors greater than a millimagnitude.

Photometric solutions for the assembled data (using the binned \wire\ light curve) and different LD laws are given in Table\,\ref{tab:lcfit}, both with and without forcing the coefficients for the two stars to be equal. Solutions with linear LD have $\chi^2_{\rm \ red} \approx 1.332$ whereas those using non-linear LD have $\chi^2_{\rm \ red} \approx 1.328$. The change in $\chi^2$ is about 20$\sigma$. According to an F-ratio test this difference is significant beyond the 99.9\% level, and we prefer solutions with non-linear LD (which have a lower $\chi^2_{\rm \ red}$) as we {\em a priori} expect these to be the most physically realistic.

It is pleasing to discover that comparable solutions with the two different non-linear LD laws are overall in very good agreement. The greatest variation between all the solutions is seen for $r_{\rm A}+r_{\rm B}$, which is clearly larger when non-linear LD is used compared to linear LD. This is qualitatively as expected (see Section\,\ref{sec:nld}). We confirm the finding of Paper\,I and \citet{Lacy+05aj} that using non-linear LD for dEBs causes the derived fractional stellar radii to be larger. Whilst \citet{Lacy+05aj} and Paper\,I found that the increase was or the order of 0.2\%, here we find it is 0.3--0.5\%. This can easily be attributed to the shallowness of the eclipses shown by \baur. We also note that including non-linear LD coefficients causes the best-fitting orbital inclination to decrease slightly, by 0.06$^\circ$. This is substantially larger than the random errors found using Monte Carlo simulations.

A comparison of the measured LD coefficients for \baur\ with those calculated using theoretical stellar atmosphere models is given in Table\,\ref{tab:ld}. We have adopted the effective temperatures and surface gravities for \baur\ from NJ94. It is possible that the values we obtain for the fitted LD coefficients are slightly affected by the {\sc ebop} approximations to the stellar shapes and the reflection effect, but this bias should be negligible. The interpretation of Table\,\ref{tab:ld} is hindered because the photometric passband of the \wire\ star tracker is not precisely known, but the theoretical predictions are generally in line with the observed coefficients. The linear LD coefficients are a little smaller than most of the predicted ones, an effect which was also seen in Paper\,I for $\psi$\,Cen and \citet{Me++04mn2} for V453\,Cyg. Further investigation would be useful to see if this is an emerging pattern, and this effect should be borne in mind when modelling the light curves of dEBs using the linear LD law.

\subsection{Error analysis}

There are two separate contributions to the uncertainties of the derived photometric parameters: statistical and systematic. Statistical errors result from the measurement errors in the data and can be assessed using Monte Carlo simulations. Systematic errors may be caused by approximations in the model used to fit the light curve and can be estimated by using a different model to fit the data independently. Here we will do both.

\subsubsection{Statistical errors}

The formal errors which are often derived from the covariance matrices of the best fit to the light curve of a dEB can significantly underestimate the true uncertainties in the solution \citep{Popper84aj,Popper+86aj,MaceroniRucinski97pasp,Me++05aa,MeClausen07aa}. This is because they do not fully take into account the correlations between different photometric parameters \citep[see][]{WilsonVanhamme03}. A good way of determining robust statistical uncertainties is by using Monte Carlo simulations \citep[][p.\,684]{Press+92book}. This has been implemented in {\sc jktebop} \citep{Me++04mn2,Me+04mn3} and found to give results which are in excellent agreement with independent parameters determined from different light curves of the same dEB \citep{Me++04mn2,Me+05mn}.

\begin{figure*} \centering \includegraphics[width=\textwidth]{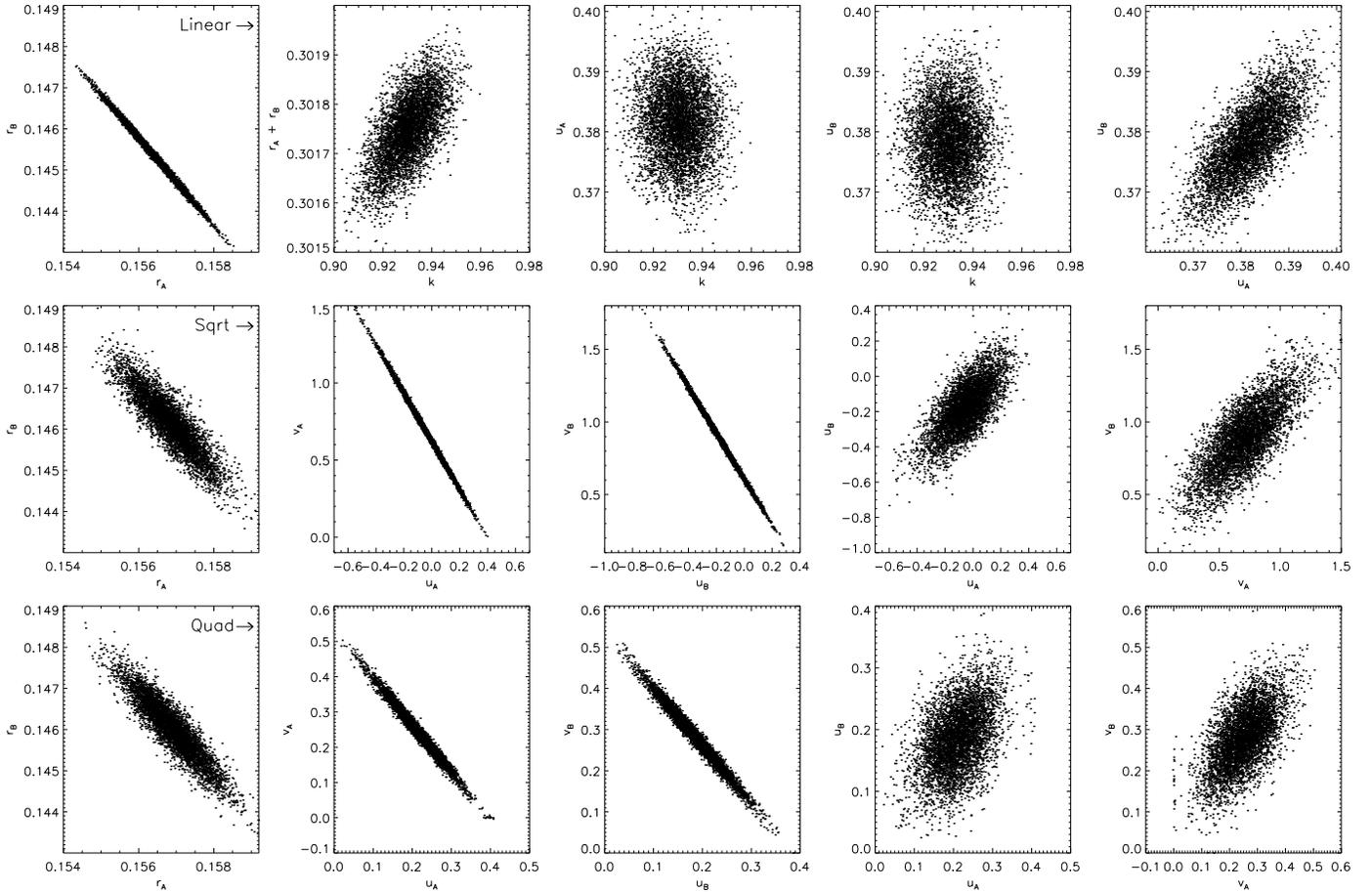}
\caption{Spreads in the values of different photometric parameters from
the Monte Carlo simulations. 5000 simulations were done for each set of
results shown here. The top row contains results for linear LD, the
middle row for square-root LD, and the bottom row for quadratic LD.}
\label{fig:MC} \end{figure*}

The uncertainties for the photometric parameters given in this work have been found using Monte Carlo simulations (5000 per fit). We have quoted 1\,$\sigma$ (68.3\%) confidence intervals for every quantity which has a value derived from each Monte Carlo simulation, except $\chi^2_{\rm \ red}$. To reduce the amount of computing time required we used \wire\ light curve after binning to reduce the number of points by a factor of six.

The results of the Monte Carlo simulations are also excellent for searching for correlations between parameters. In this case the light curve is generally very well defined except for the ratio of the radii, $k$, as previously discussed. In Fig.\,\ref{fig:MC} are shown plots of the spreads of values of various parameters, concentrating on those which are important or otherwise interesting. The poorly defined $k$ means that $r_{\rm A}$ and $r_{\rm B}$ are quite correlated -- as $r_{\rm A} + r_{\rm B}$ is well defined their values are very strongly dependent on the spectroscopic light ratio.

The LD coefficients are well defined in the case of the linear law, but it is clear that for the non-linear laws the two coefficients for each star are hugely correlated, which is not surprising. This seems to be worse for the square-root law than for the quadratic law, although this does not cause the overall light curve solutions for the two laws to be different either in absolute value or uncertainty. A much weaker correlation is seen between comparable coefficients for the two stars. This shows that in general non-linear LD coefficients can only be measured using light curves of a comparable quality to the \wire\ observations of \baur. However, we would expect that dEBs with much deeper or total eclipses would give comparatively better results. A small clustering of points for the Monte Carlo simulations using quadratic LD is visible at $v_{\rm A} \approx 0.0$. This is due to a slight numerical problem when calculating symmetrical partial derivatives. It is sufficiently far from the best-fitting value to have no effect except to slightly increase the derived uncertainties.

\subsubsection{Systematic errors}

{\sc ebop} is suitable only for the analysis of well-detached eclipsing binaries because it uses quite a simple approximation to the shapes of deformed stars. This approximation may be checked by fitting a light curve with a more complex code which uses Roche geometry to approximate the surfaces of the stars. We have therefore studied the \wire\ dataset using the Wilson-Devinney code \citep{WilsonDevinney71apj,WilsonDevinney73apj,Wilson79apj} in its 2003 version (hereafter {\sc wd2003}).

As {\sc wd2003} does not have the option of directly fitting times of minimum light, we chose to consider only the (binned) \wire\ data with the orbital ephemeris fixed to that obtained during our {\sc jktebop} analysis. The square-root LD law was used and the coefficients were interpolated from the tabulations by \citet{Vanhamme93aj} assuming a Cousins $R$ passband. The linear coefficients were included as fitted parameters but the non-linear ones were not. The effective temperatures were fixed at the values given by NJ94 and the {\sc wd2003} mode 0 was used so the light contributions of the two stars were not connected to these quantities. To force the solution to reproduce the spectroscopic light ratio from NJ94 we fixed the light contribution from star A to an appropriate value. This rather brute-force approach was needed because {\sc wd2003} does not have a facility to fix the light ratio directly. Due to this approach, the formal errors of the fitted parameters (which come from the covariance matrix) are unrealistic so we will not quote them.

\begin{table} \caption{Parameters of our solution of the \wire\ light curve
using {\sc wd2003}. The optimised parameters are given in bold type and the
fixed parameters in normal type. The rotation rates of the stars are given
in units of the synchronous velocity.
\newline {\bf References:} (1) \citet{Smith48apj}; (2) NJ94; (3) \citet{Claret98aas}.}
\label{tab:wdfit} \centering
\begin{tabular}{l c c r} \hline \hline
Parameter                     & Star A          & Star B          & Ref.\\
\hline
Integration accuracy (points) & 60              & 60              &     \\
Time of primary min. (HJD)    & \mc{2453827.19569}                &     \\
Orbital period (day)          & \mc{3.9600467300}                 &     \\
Phase shift                   & \mc{\bf $-$0.000001}              &     \\
Mass ratio                    & \mc{0.964}                        & 1   \\
Orbital inclination           & \mc{\bf 76.909}                   &     \\
Orbital eccentricity          & \mc{\bf 0.00203}                  &     \\
Longitude of periastron (deg) & \mc{\bf 90.30}                    &     \\
Mean efective temperatures (K)& 9350            & 9200            & 2   \\
Potentials                    & {\bf 7.3309}    & {\bf 7.5704}    &     \\
Rotation rates                & 1.0             & 1.0             &     \\
Bolometric albedo             & {\bf 0.590}     & {\bf 0.558}     &     \\
Gravity brightening exp.      & 1.0             & 1.0             & 3   \\
$u$                           & {\bf 0.109}     & {\bf 0.102}     &     \\
$v$                           & 0.624           & 0.624           &     \\
Light output                  & 6.775           & {\bf 5.801}     &     \\
Third light                   & \mc{0.0}                          &     \\
\hline
$r$ (volume)                  & 0.15717         & 0.14550         &     \\
$r$ (point)                   & 0.15648         & 0.14495         &     \\
$r$ (pole)                    & 0.15807         & 0.14621         &     \\
$r$ (side)                    & 0.15707         & 0.14541         &     \\
$r$ (back)                    & 0.15787         & 0.14607         &     \\
\hline \end{tabular} \end{table}

During the initial solutions (with an integration accuracy of $N1 = N2 = 30$ points; \citealt{WilsonVanhamme03}) it was not possible to get a good fit to the light curve. The problem arose because the amplitude of the light variation outside eclipses was being underestimated, resulting in the profile of the eclipse being very poorly matched. We were able to solve the problem by using the maximum integration accuracy ($N1 = N2 = 60$), including the bolometric albedos as fitting parameters, and executing a large number of iterations. Due to the high integration accuracy and the eccentric orbits required to fit \baur, using the detailed reflection effect in {\sc wd2003} would have been prohibitively expensive in terms of calculation time. We therefore used the simple reflection effect, but preliminary analyses indicated that this makes a negligible difference to our results. Our final {\sc wd2003} solution has slightly larger residuals than for the {\sc jktebop} solution, but the preliminary analyses have shown that this does not have a significant effect on the final parameters. Because of this our {\sc wd2003} analysis should not be regarded as definitive, and we hope to improve it by implementing a more modern minimisation algorithm into {\sc wd2003} (see discussion in Paper\,I).

The final {\sc wd2003} solution is given in Table\,\ref{tab:wdfit} and is in extremely good agreement with the {\sc jktebop} results. The greatest difference between the {\sc jktebop} and {\sc wd2003} solutions is for the orbital inclination, which is 0.1$^\circ$ ($4.6 \sigma$) higher in {\sc wd2003} than {\sc ebop}. A similar result has been found by \citet{PopperEtzel81aj} and \citet{Andersen++93aa} when comparing {\sc ebop} and {\sc wink} \citep{Wood71aj,Wood73pasp}, and attributed to the treatment of the stellar shapes in {\sc ebop}. However, the size of this effect is not enough to cause significant problems. This systematic error has been included in further analyses. For the other parameters there is no reason to believe that systematic errors are significant so we have retained their statistical errors as the final uncertainties.

\reff{In the {\sc wd2003} analysis the bolometric albedos (Table\,\ref{tab:wdfit}) converged to values near the canonical 0.5 for convective atmospheres rather than the expected 1.0 for radiative atmospheres \citep[see][]{RafertTwigg80mn,Claret01mn}. This is surprising because the effective temperatures of the components of \baur, 9350\,K and 9200\,K (NJ94) are appropriate for radiative atmospheres \citep{Claret00aa2,Hurley++02mn}. We are also not aware of any similar cases in the literature. However, as we do not have realistic uncertainties for the best-fitting bolometric albdeo values (see above), it is not clear how significant this discrepancy is. Further investigations of this effect will be undertaken in the future.}

\subsection{Final light curve parameters}

\begin{table} \caption{Final photometric parameters for \baur\ from fitting the \wire\
and Hipparcos light curves, the spectroscopic light ratio of NJ94 and several times of
minimum light. The quadratic LD law was used with independently-optimised LD coefficients
for the two stars. The upper part of the table contains directly optimised parameters and
the lower part contains dependent parameters. The uncertainties have been multiplied by
$\sqrt{\chi^2_{\rm \ red}}$.}
\label{tab:lcfinal} \centering
\begin{tabular}{l l r@{\,$\pm$\,}l} \hline \hline
Parameter                     & Symbol                      & \mc{Value}                  \\
\hline
Sum of relative radii         & $r_{\rm A}+r_{\rm B}$       &      0.30290    & 0.00054   \\
Ratio of radii                & $k$                         &      0.9300     & 0.0097    \\
Orbital inclination (deg.)    & $i$                         &     76.80       & 0.10      \\
Eccentricity term             & $e\cos\omega$               &   $-$0.00002    & 0.00001   \\
Eccentricity term             & $e\sin\omega$               &      0.00183    & 0.00040   \\
Linear LD coefficient         & $u_{\rm A}$                 &      0.210      & 0.074     \\
Linear LD coefficient         & $u_{\rm B}$                 &      0.187      & 0.069     \\
Non-linear LD coefficient     & $v_{\rm A}$                 &      0.248      & 0.111     \\
Non-linear LD coefficient     & $v_{\rm B}$                 &      0.277      & 0.108     \\
\hline
Relative radius of star A     & $r_{\rm A}$                 &      0.15694    & 0.00081   \\
Relative radius of star B     & $r_{\rm B}$                 &      0.14595    & 0.00082   \\
Light ratio                   & $\ell_{\rm B}/\ell_{\rm A}$ &      0.854      & 0.016     \\
Orbital eccentricity          & $e$                         &      0.00183    & 0.00040   \\
Longitude of periastron (deg) & $\omega$                    &     90.47       & 0.26      \\
\hline \end{tabular} \end{table}

It is very pleasing to discover that the six solutions listed in Table\,\ref{tab:lcfit} give very similar results. The two with non-linear LD and separate coefficients for the two stars have the lowest $\chi^2_{\rm \ red}$. Of these two the quadratic LD law solution has a marginally better $\chi^2_{\rm \ red}$ and gives the best-defined LD coefficients, so we will adopt this as our final solution. The two solutions differ by less than 0.1\% in $r_{\rm A}$ and $r_{\rm B}$. The quadratic-LD fit is shown in Fig.\,\ref{fig:lcfit} and the times of minimum light are given in Table\,\ref{tab:tmin}. We find that the ephemeris of the system is
\[ T_{\rm Min~I} ({\rm HJD}) = 2\,452\,827.195693 (12) + 3.96004673 (17) E \]
where $E$ is the cycle number and the quantities in parentheses indicate the uncertainty in the final digit of the preceding number. The quoted uncertainty for the constant term corresponds to a probably optimistic 1.0\,s, which is of a similar magnitude to small systematic effects such as the difference between heliocentric and barycentric Julian Date.

The uncertainties in Table\,\ref{tab:lcfinal} come from the Monte Carlo simulations and have been multiplied by $\sqrt{\chi^2_{\rm \ red}} = 1.152$ to account for the small systematic errors detected in Section\,\ref{sec:lc:soln}. A minor systematic error was found in the orbital inclination, and its uncertainty in Table\,\ref{tab:lcfinal} has been increased to reflect this.

The final photometric parameters are more accurate than and in good agreement with those of NJ94. \citet{Hummel+95aj} found an orbital inclination of $i = 76.0^\circ \pm 0.4^\circ$ from interferometric observations -- this is 0.8$^\circ$ ($2.0\sigma$) lower than our result.


\section{Absolute dimensions and comparison with theoretical models}

\begin{table}
\caption{The physical parameters of the component stars of \baur\ derived
from the {\sc jktebop} analysis and the circular spectroscopic orbit of
\citet{Smith48apj}. The effective temperatures are from NJ94.
\newline $*$ Calculated assuming $\lsun = 3.826${$\times$}10$^{26}$\,W
and $\Mbol\sun = 4.75$ \citep{Zombeck90book}.}
\label{tab:absdim} \centering
\begin{tabular}{l r@{\,$\pm$\,}l r@{\,$\pm$\,}l} \hline \hline
Parameter                     &    \mc{Star A}     &    \mc{Star B}     \\ \hline
Orbital separation (\Rsun)    &  \multicolumn{4}{c}{$17.60 \pm  0.06$}  \\
Mass (\msun)                  &   2.376  &  0.027  &   2.291  &  0.027  \\
Radius (\rsun)                &   2.762  &  0.017  &   2.568  &  0.017  \\
$\log (g/{\rm cm\,s^{-2}})$   &   3.932  &  0.005  &   3.979  &  0.005  \\
\Vsync\ (\kms)                &   35.3   &  0.2    &   32.8   &  0.2    \\
$\log\Teff$ (K)               &   3.971  &  0.009  &   3.964  &  0.009  \\
$\log(L/\lsun)$ $*$           &   1.721  &  0.038  &   1.630  &  0.038  \\
\Mbol\ $*$                    &   0.448  &  0.094  &   0.676  &  0.095  \\
\hline \end{tabular}\end{table}

\begin{table*}
\caption{Different measurements of the distance to \baur. Details and references are given in the text.}
\label{tab:dist} \centering
\begin{tabular}{l r@{\,$\pm$\,}l r@{\,$\pm$\,}l r@{\,$\pm$\,}l r@{\,$\pm$\,}l} \hline \hline
Method                                          & \multicolumn{8}{c}{Distance (pc)}                 \\ \hline
Hipparcos parallax                              & \multicolumn{8}{c}{$25.2 \pm 0.5$}                \\
Orbital parallax                                & \multicolumn{8}{c}{$24.8 \pm 0.8$}                \\[2pt] \hline
                                                &  \mc{$B$}  &  \mc{$V$}  &  \mc{$J$}  &  \mc{$K$}  \\ \hline
Surface brightness calibrations                 & 25.5 & 1.8 & 24.8 & 1.6 & 24.8 & 0.5 & 25.0 & 0.4 \\
Bolometric corrections (Bessel et al.)          &    \mc{}   & 24.7 & 0.6 &    \mc{}   & 24.8 & 0.3 \\
Bolometric corrections (Girardi et al.)         & 25.5 & 0.8 & 24.7 & 0.6 & 24.5 & 0.4 & 24.8 & 0.3 \\
Empirical bolometric corrections (Code et al.)  &    \mc{}   & 24.4 & 0.8 &    \mc{}   &    \mc{}   \\
Empirical bolometric corrections (Flower)       &    \mc{}   & 24.9 & 0.9 &    \mc{}   &    \mc{}   \\
\hline \end{tabular}\end{table*}

The fundamental physical properties of the components of \baur\ (Table\,\ref{tab:absdim}) have been calculated from the final results of the {\sc jktebop} analysis (Table\,\ref{tab:lcfinal}), the circular spectroscopic orbit given by \citet{Smith48apj}\footnote{\citet{Smith48apj} gives probable errors for the orbital elements he derives. These must be multiplied by a factor of 1.48 to give standard errors \citep[e.g.][]{Griffin05obs}, so the velocity amplitudes used here are $K_{\rm A} = 107.46 \pm 0.58$\kms\ and $K_{\rm B} = 111.49 \pm 0.55$\kms.}, and the effective temperatures given by NJ94. The masses are known to an accuracy of 1.2\% and the radii to 0.5\%, and the stars are slightly evolved with surface gravities of just below 4.0 (cgs units).

The distance to \baur\ can be calculated in several ways (Table\,\ref{tab:dist}). The most direct of these are the trigonometric parallax measurements from Hipparcos \citep{Perryman+97aa}, and the orbital parallax obtained from the linear separation of the stars (Table\,\ref{tab:absdim}) and the orbital angular size from \citet{Hummel+95aj}. These two measurements are in excellent agreement.

The distance to \baur\ can also be calculated using surface brightness calibrations or bolometric corrections \citep{Me++05aa}, both of which require reliable apparent magnitudes. We have corrected the mean Hipparcos $B_T$ and $V_T$ magnitudes to remove the effect of points within eclipse, and supplemented these with $JK$ magnitudes from \citet{Johnson66} adopting estimated uncertainties of 0.02\,mag for the $JK$ data. Applying the calibration of \citet{MoonDworetsky85mn} to the Str\"omgren indices given by NJ94, we find that the reddening is negligible. This is unsurprising given the closeness of \baur\ to the Sun (25\,pc) so we adopt $\EBV = 0.00 \pm 0.01$.

We have estimated the distance to \baur\ by the surface brightness method\footnote{{\sc fortran77} code for calculating the absolute dimensions and various distance measurements for a dEB is available at {\tt http://www.astro.keele.ac.uk/$\sim$jkt/}} presented by \citet{Me++05aa}, which uses calibrations of surface brightness against effective temperature \citep{Kervella+06aa}. We have also calculated the distance using theoretical bolometric corrections \citep[see][]{Me++05iauc} from \citet{Bessell++98aa} and \citet{Girardi+02aa} and empirical bolometric corrections from \citet{Code+76apj} and \citet{Flower96apj}. The good news is that the agreement between all four different methods of distance determination is excellent. Whilst the two parallax distances are the most reliable because they are empirical direct measurements, formally the most accurate results are obtained using bolometric corrections or surface brightness relations in the $K$ band, where uncertainties in effective temperature and interstellar extinction have the smallest effect.

We have compared the predictions of several sets of theoretical stellar evolutionary models to the properties of \baur. These sets are the Granada models \citep{Claret95aas,Claret97aas,ClaretGimenez95aas}, the Padova models \citep{Girardi+00aas}, and the Cambridge models \citep{Pols98mn}. In each case a good fit can be found to the masses, radii and effective temperatures of \baur\ for an approximately solar metal abundance ($Z = 0.02$) and an age of between 450 and 500 Myr. For a change in $Z$ of 0.01 the quality of the fit is slightly decreased and the age changes by about 50\,Myr, with a higher $Z$ requiring a lower age. The properties of the two stars are too similar for a detailed comparison with theoretical predictions to be useful given the accuracy of the absolute dimensions we have established. An improved spectroscopic orbit will be needed before further conclusions can be made.


\section{Discussion and conclusions}

In this work we have presented a high-quality light curve of the detached eclipsing binary system \baur igae, containing 30\,015 datapoints with a point-to-point scatter of less than 0.3\,mmag, obtained using the star tracker on board the \wire\ satellite. Light curves of this quality will become available in the future for dEBs (from space missions such as Kepler and CoRoT), and it is important to ensure that the accuracy of light curve models will not simply be overwhelmed by these datasets.

The data have been analysed using a version of the {\sc ebop} code which has been extended to include non-linear limb darkening laws and the direct inclusion of observed times of minimum light and spectroscopic light ratios in the light curve model fit. Robust uncertainties have been obtained using extensive Monte Carlo simulations. The eclipses exhibited by \baur\ are partial and quite shallow, so the ratio of the stellar radii is poorly defined by the photometric data. Including the accurate spectroscopic light ratio obtained by NJ94 causes the uncertainties in the stellar radii to drop by a remarkable amount, from 5\% to 0.5\%.

The importance of including non-linear limb darkening laws in the analyses of dEB light curve has been investigated using the quadratic and square-root laws. An F-ratio test indicates that the resulting improvement in the fit is highly significant, and we find that the measured radii of the stars are 0.4\% greater when using non-linear LD compared to the simple linear law, in agreement with previous studies. This bias is large enough to be important and we recommend that non-linear LD laws are used in photometric analyses of the light curves of dEBs. \baur\ is not the best system to test this because its eclipses are shallow, so it is possible that similar conclusions could be reached using lower-quality data of a deeply or totally eclipsing system.

The LD coefficients obtained for \baur\ for the linear, square-root and quadratic laws are in reasonable agreement with theoretical predictions from model atmosphere calculations. We find that the coefficients of bi-parametric laws are strongly correlated with each other, and that this will prevent their simultaneous measurement except for datasets of comparable quality to the \wire\ light curve. The quadratic LD coefficients are better determined than the square-root coefficients, which suggests that the latter law may actully be preferable as changes in the coefficients have less effect on the other parameters of the fit. The light curve fits for all three LD laws are in good agreement, so the choice of non-linear law is not important.

In order to detect any systematic errors in the {\sc jktebop} fit to the data, we have also fitted the \wire\ light curve using the 2003 version of the Wilson-Devinney code. We were able to obtain a reasonable fit but only by using the highest available integration accuracy, bolometric albedos appropriate to convective rather than radiative envelopes, and a large number of iterations. The results agree well with the {\sc ebop} solution, expect for a difference of 0.1$^\circ$ in orbital inclination, which indicates that any systematic errors are too small to be important.

The physical properties of the component stars of \baur\ have been calculated from the results of the {\sc jktebop} analysis and the spectroscopic orbit of \citet{Smith48apj}, resulting in radii known to an accuracy of about 0.5\% and masses to an accuracy of 1.2\%. A good agreement between these properties and the predictions of theoretical stellar evolutionary models is found for an approximately solar chemical composition and an age of 450 to 500\,Myr. This comparison is strongly dependent on the derived ratio of the stellar radii, which is itself determined primarily by the spectroscopic light ratio from NJ94. We will be obtaining new CCD spectroscopic observations of \baur\ in order to calculate a more accurate spectroscopic orbit and light ratio, with the aim of measuring the masses as well as the radii of the components of \baur\ to accuracies of 0.5\%.

This analysis has shown that {\sc jktebop} can provide an excellent fit to the high-precision \wire\ light curve of the bright eclipsing system \baur. Solutions using {\sc wd2003} indicated that there were no significant systematic effects in the {\sc jktebop} solution. {\sc wd2003} was able to give a reasonable fit the the \wire\ data, but at the expense of a large amount of computation time. We should therefore be able to derive highly reliable photometric solutions for the expected deluge of space-based eclipsing binary light curves from the Kepler and CoRoT missions.


\section*{Acknowledgements}

JS acknowledges financial support from PPARC in the form of a postdoctoral research assistant position. JS is grateful to Jens Viggo Clausen, Bob Wilson, Tom Marsh and Antonio Claret for wide-ranging discussions. We also thank the referee for a prompt report and useful comments.

The following internet-based resources were used in research for this paper: the NASA Astrophysics Data System; the SIMBAD database operated at CDS, Strasbourg, France; the ar$\chi$iv scientific paper preprint service operated by Cornell University; and the General Catalogue of Photometric Data maintained by J.-C.\ Mermilliod, B.\ Hauck \& M.\ Mermilliod at the University of Lausanne, Switzerland.


\bibliographystyle{aa}



\end{document}